\documentstyle[11pt,cite,amssymb,url,color,graphicx,fontenc]{article}

\begin{document}

\begin{center}
{\LARGE 

\textbf{Reflection efficiency and spectra resolutions ray-tracing simulations for the VOXES HAPG crystal based Von Hamos spectrometer}}
\end{center}

\vspace{0.5cm}

\textsc{V. De Leo$^a$, A. Scordo$^b$, C. Curceanu$^b$, M. Miliucci$^b$, F. Sirghi$^b$}


\vspace{0.5cm}
\begin{center}
\small
$^a$Universit\`a degli Studi di Roma "La Sapienza", INFN Sezione Roma 1, Rome, Italy \\
$^b$Laboratori Nazionali di Frascati, Frascati (Rome) Italy
\end{center}

%
%





\vspace{0.2cm}
\abstract{
The VOXES collaboration at INFN National Laboratories of Frascati developed a prototype of a high resolution Von Hamos X-ray spectrometer using HAPG (Highly Annealed Pyrolytic Graphite) mosaic crystals. This technology allows the employment of extended isotropic sources and could find application in several physics fields.
The capability of the spectrometer to reach energy precision and resolution below 1 and 10 eV, respectively, when used with wide sources have been already demonstrated. 
Recently, the response of this device, for a $\rho$ = 206.7 mm cylindrically bent HAPG crystal using $CuK\alpha_{1,2}$ and $FeK\alpha_{1,2}$ XRF lines, has been investigated in terms of reflection efficiency by a dedicated ray-tracing simulation.
Details of the simulation procedure and the comparison with the experimental results are presented. This study is crucial in order to retrieve information on the  spectrometer signal collection efficiency, especially in the energy range in which the standard calibration procedures cannot be applied.
}

\section{Introduction}
One of the most quoted methods to perform high energy resolution X-ray measurements is provided by the Bragg spectroscopy. The requirement on the size of the target not to exceed tens of microns represents the major hindrance in its use when  photons emitted from extended sources (millimetric) need to be measured \cite{Barn}. In addition, the typical very low efficiencies of Bragg spectrometers prevent them from being used in several applications.
\noindent The prototype of a high resolution Von Hamos X-ray spectrometer using HAPG (Highly Annealed Pyrolytic Graphite) mosaic crystals developed by the VOXES collaboration at INFN National Laboratories of Frascati offers the possibility to achieve few eV energy resolution for energies going from 2 keV up to 10 of keV and to measure not only collimated sources, but also extended ones. Highly Annealed Pyrolitic Graphite (HAPG) is a mosaic crystal consisting in a large number of nearly perfect small crystallites. Mosaicity makes it possible that a photon can find a crystallite plane at the right Bragg angle and be reflected although if it is slightly deviated from reaching the crystal with the exact Bragg relation \cite{Legall,Grigorieva}. 
This property together with a lattice spacing constant
$d=3.514$ °A, enables them to be highly efficient in the 2–20 keV energy range. This technology is suitable to be used in the Von Hamos configuration combining the standard dispersion of a flat crystal with the focusing properties of cilindrically bent crystals.

\noindent Several studies aimed to analyze the effect on the energy resolution, of the precision of the mosaicity and thickness of the crystal have been already performed and the capability of the spectrometer to be optimized in order to achieve the best precision has been also demonstrated \cite{Scordo_Cond,Scordo_JAAS}. 

\noindent The performance of the spectrometer has been lately studied in terms of reflection efficiency \cite{nostro_JAAS} and in this work we want to focus on the ray-tracing simulations implemented in order to check the agreement with the experimental results. The achievement of consistent results is fundamental to evaluate the efficiency of the spectrometer.     
In the next sections the description of the experimental setup and procedure, of the ray-tracing simulations, and the comparison of the experimental and simulated data are presented. 
\section{Setup}
In the spectrometer configuration used in our measurements the X-ray source and the position detector are placed on the axis of a cylindrical crystal. This configuration is known as Von Hamos and allows to increase the reflection efficiency due to the vertical focusing. This geometry permits to determine the source-crystal and the source-detector distances, ${L_1}$ and $L_2$ respectively, by means of the Bragg angle ($\theta_B$) and of the curvature radius of the crystal ($\rho_c$):

\begin{equation}
L_1=\frac{\rho_c}{sin_{\theta_B}}
\end{equation}
\begin{equation}
L_2=L1\, sin\phi
\end{equation}

\noindent where $\phi$=$\pi$-$\theta_B$.


\begin{figure}
    \centering
    \includegraphics[width=0.6\linewidth]{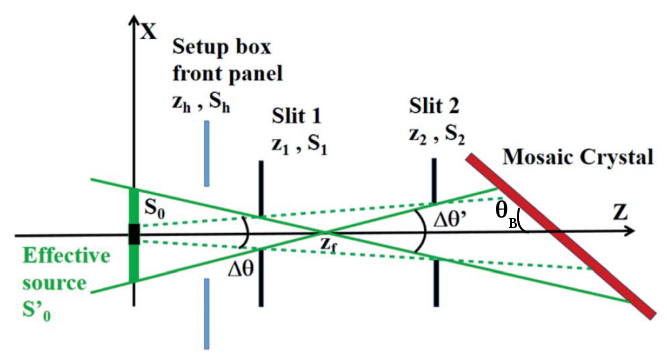}
    \caption{XRF beam dispersive plane (not in scale) where is represented the slits system ($S_1$ and $S_1$), the virtual and effective sources, $S_0$ and  $S_0'$, the angular acceptance $\Delta\theta'$, the Bragg angle $\theta_B$ and the mosaic crystal (see \cite{Scordo_Cond} for more details).}
    \label{Horizontal_vertical plane}
\end{figure}

\noindent Adding a pair of slits to this configuration, it is possible to shape the beam of the X-rays emitted by an extended target modifying the position ($Z_1$ and $Z_2$) and the apertures ($S_1$ and $S_2$) of the slits to create a virtual point-like source ($Z_f$), an angular acceptance $\Delta\theta'$ and an effective source $S_0'$ (see Fig.~\ref{Horizontal_vertical plane} for the horizontal plane of the beam)\cite{Scordo_Cond}.



\noindent The experimental apparatus consists of a XTF-5011 Tungsten anode X-ray tube produced by OXFORD INSTRUMENTS, located on the top of an aluminum box where a 125 $\mu$m thick target foil is contained. 
The center of the foil is the source position and it is placed on a 45$^\circ$ target holder. 

\noindent The two adjustable motorized slits (STANDA 10AOS10-1) with 1 mm sensitivity
are arranged after the circular exit window of the aluminum box of 5.9 mm diameter.
The HAPG crystal used for the measurements has
a thickness of 100 $\mu$m, a curvature radius ($\rho_c$) of 206.7 mm and mosaicity ($\omega$) of 0.1$^\circ$. 
The position detector, also equipped with a positioning motorized
system, is a commercial MYTHEN2-R-1D 640
channels strip detector produced by DECTRIS (Zurich, Switzerland). 
The active area is 32 $\times$ 8 mm$^2$, strip width and thickness are, respectively, 50 $\mu$m and 420 $\mu$m; further details can be found in \cite{Scordo_JAAS}.
\section{Experimental procedure}
\begin{figure}[h!]
    \centering
    \includegraphics[width=0.7\linewidth]
    {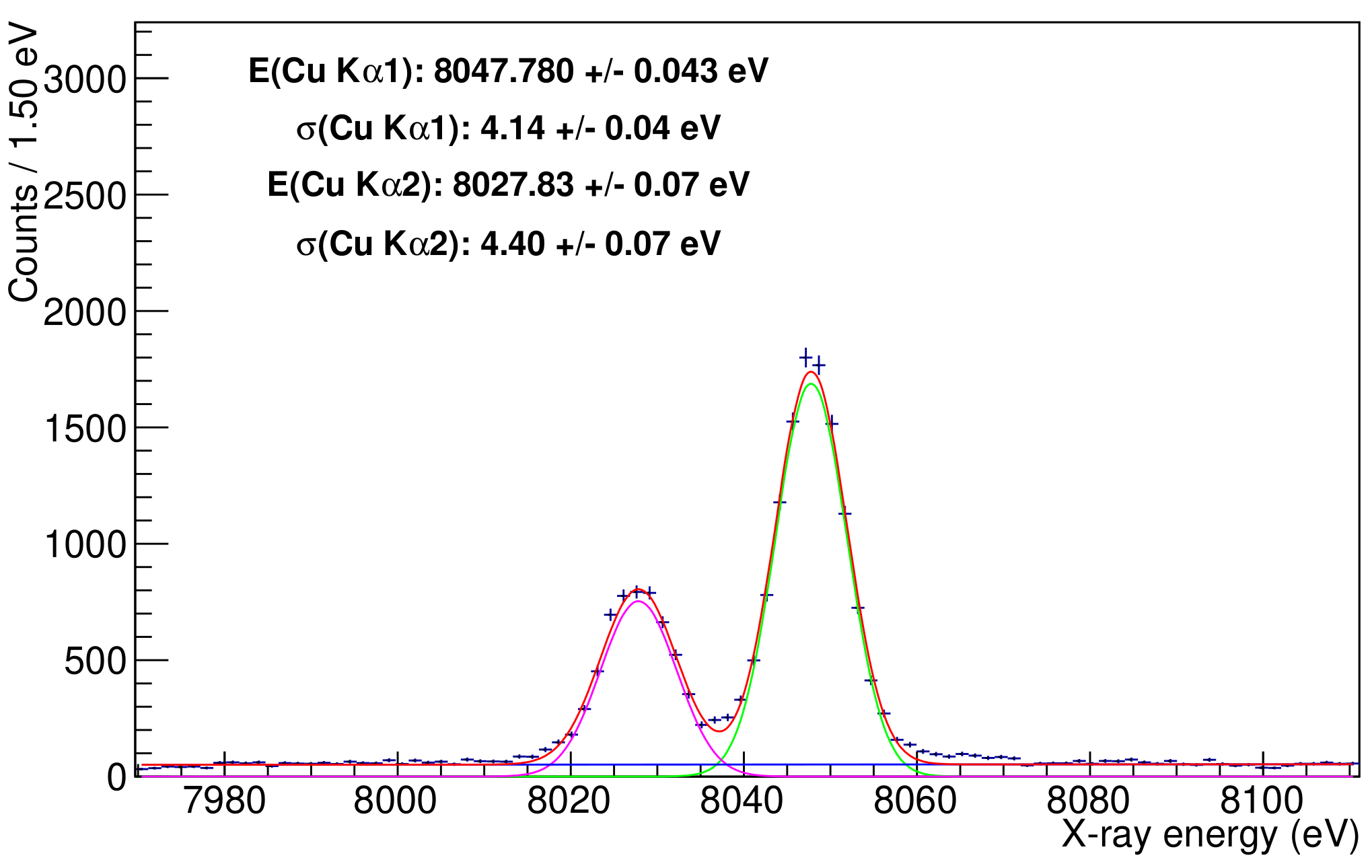}
    \caption{Copper Bragg spectrum
    where the overall fitting function (red), the polynomial background (blue) and the $K\alpha_1$ (green) and $K\alpha_2$ (violet) components are over imposed to the spectrum for $\Delta\theta$=0.7 and $S_0'$=1.1 mm, elaborated from \cite{Scordo_JAAS} (see text for details).}
    \label{exp_spectrum}
\end{figure}

The spectrometer setup was optimized for the measurement of the two Cu and Fe $K\alpha$ lines. The target foil is activated by the X-ray tube and the $K\alpha_{1,2}$ lines, isotropically emitted, are collimated by the slits system to simulate a point-like source. An example of the resulting spectrum, after performing the calibration and fitting procedure reported in our previous work \cite{Scordo_JAAS}, is shown in the Fig.~\ref{exp_spectrum}. In this case the slits are set to maintain an angular divergence $\Delta\theta$ of 0.7$^\circ$ and the distances between the HAPG source-to-crystal and crystal-to-detector are of 900.54 mm. The alignment of the optical system has been performed by using a laser.  

\noindent These kind of spectra are used in the evaluation of the HAPG crystal reflection efficiency.
Usually this quantity is defined as the percentage of Bragg reflected X-rays of a given energy $E_0$ for different impinging angles and is obtained from X-ray beam emitted from a monochromatic point-like source. 
Instead, the quantity we want to measure is the following:

\begin{equation}\label{effr_eq}
\epsilon^R_{\Delta\theta',S_0'} = \frac{R^r_{\Delta\theta',S_0'}}{R^i_{\Delta\theta',S_0'}}
\end{equation}

\begin{figure}[h!]
    \centering
    \includegraphics[width=0.8\linewidth]
    {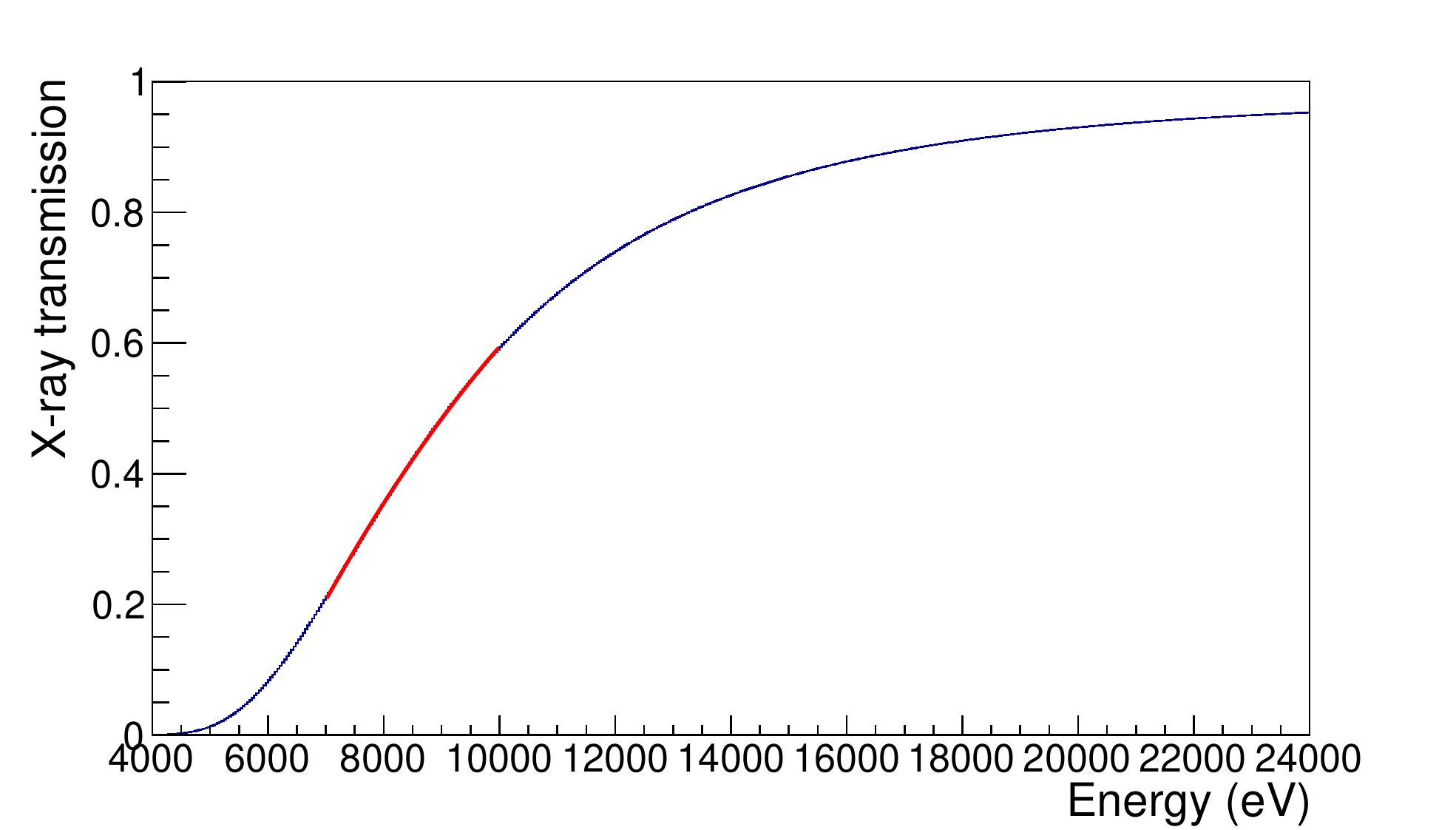}
    \caption{The X-ray transmission coefficient function of energy; the performed fit in the energy range of our interest is represented in red.} 
    \label{transmission}
\end{figure}

\noindent It represents the ratio between the number of $Cu(K_{\alpha1,2})$ or $Fe(K_{\alpha1,2})$ X-rays reflected from the crystal ($R^r_{\Delta\theta',S_0'}$) and that of those impinging on it ($R^i_{\Delta\theta',S_0'}$) 
for different source sizes $S_0'$ and beam divergence $\Delta\theta'$ pairs \cite{nostro_JAAS}.
The numerator of Eq. \ref{effr_eq} is obtained from the Bragg energy spectrum $R^B_{\Delta\theta',S_0'}$, as the one shown in Fig.~\ref{exp_spectrum}, taking into account the X-ray transmission in air ($T_{air}$) and the MYTHEN2 detection efficiency ($QE^M$):

\begin{equation}\label{effr_num}
R^r_{\Delta\theta',S_0'} = R^B_{\Delta\theta',S_0'}\frac{1}{T_{air}}\frac{1}{QE^M}
\end{equation}

\noindent The $T_{air}$ coefficient is evaluated, for a given energy and HAPG-MYTHEN2 distance, from the CXRO database \cite{CXRO} while $QE^M$ is provided by the producers. 
In the Fig.~\ref{transmission} the resulting behaviour of the transmission coefficient with the energy is shown; the fit carried out in the energy range of interest for our measurements is represented in red.

\begin{figure}[h!]
    \centering
    \includegraphics[width=0.6\linewidth]
    {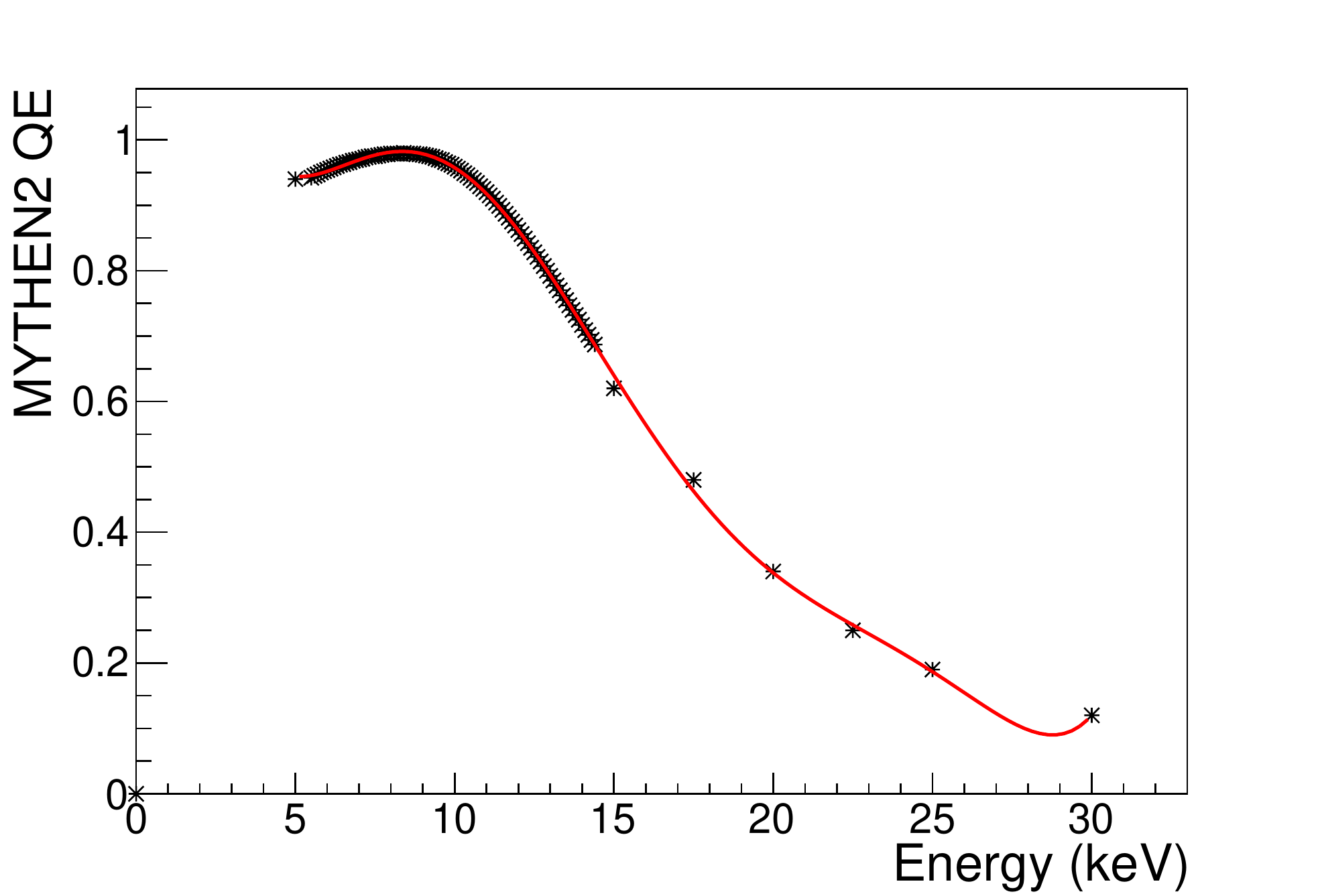}
     \includegraphics[width=0.6\linewidth]
    {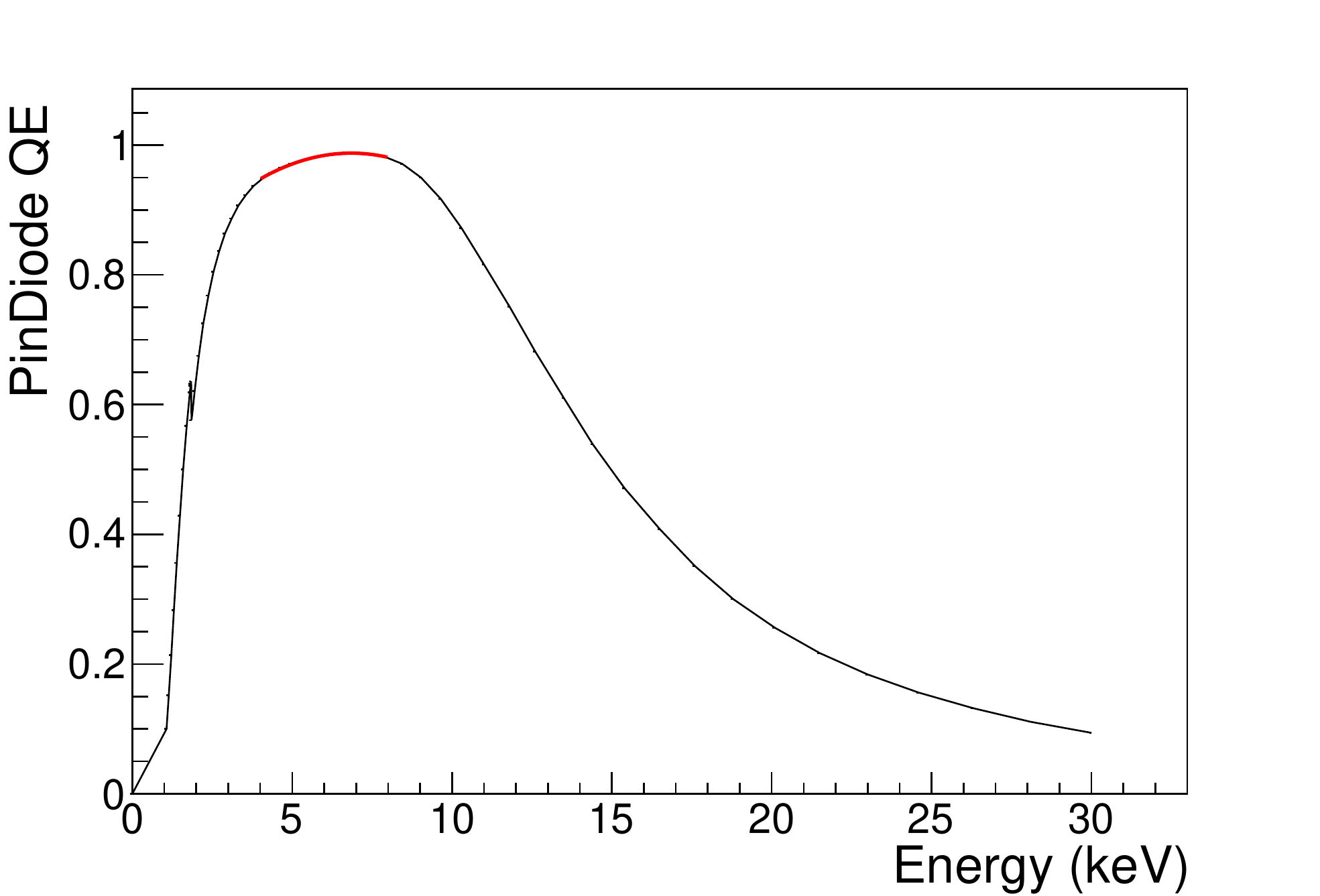}
    \caption{MYTHEN2 detector (top) and  PinDiode quantum efficiencies (bottom) as a function of energy.}
    \label{bin_correction}
\end{figure}

\noindent The determination of the number of X-rays impinging on the crystal ($R^i_{\Delta\theta',S_0'}$) required a longer procedure. 
First of all, for each $S_0'\,\Delta\theta'$ combination, a measurement with the MYTHEN2 detector in place of the HAPG crystal has been performed in order to quantify the X-rays reaching the HAPG position that we defined as MYTHEN2 direct measurements $R^M_{\Delta\theta',S_0'}$.
However, since with the MYTHEN2 detector it was not possible to make the energy selection  and we have to find the numbers of pure $Cu(K_{\alpha1,2})$ or $Fe(K_{\alpha1,2})$ signal,
we made use of a Silicon PinDiode to extract the ratio $R(S/B)$ between the $Cu(K_{\alpha1,2})$ or $Fe(K_{\alpha1,2})$ lines and all the other background photons, produced by Bremmstrahlung or other processes occurring in the source box and reaching the crystal.
The spectrum is bin by bin corrected accounting for the different efficiencies of the PinDiode and of the MYTHEN2 detector (see Fig.~\ref{bin_correction}):
\begin{equation}
Bin_{PD_{corr}} = Bin_{PD} \times QE_{M}/QE_{PD}
\end{equation}

\noindent The number of X-rays impinging on the HAPG crystal can then be evaluated from the following relation:

\begin{equation}\label{effr_den}
R^i_{\Delta\theta',S_0'} = R^M_{\Delta\theta',S_0'}\frac{1}{QE^M}R(S/B)\frac{A_c}{8\,mm}
\end{equation}

\noindent where the ratio between the beam vertical dimension $A_c$ and $8\,mm$ accounts for the difference between the MTYTHEN2 strip height and the XRF beam spot vertical size on the HAPG. $A_c$ is evaluated, following the procedure described in \cite{Scordo_JAAS}, from the slits positions and dimensions. 


\noindent The reflection efficiencies, obtained with a $\rho = 206.7 \,mm$ HAPG crystal for both $Cu_{K\alpha 1,2}$ $Fe_{K\alpha 1,2}$ XRF lines, from Eq. \ref{effr_num} and \ref{effr_den}, are estimated as\cite{nostro_JAAS}:

\begin{equation}\label{effr_expl}
\epsilon^R_{\Delta\theta',S_0'} = \frac{8 R^B_{\Delta\theta',S_0'}}{T_{air} R^M_{\Delta\theta',S_0'}R(S/B){A_c}}
\end{equation}
 
\noindent and resulted to vary between 0.15-0.35 and 0.1-0.3 for copper and iron, respectively.



    

\section{Ray-tracing Simulations}
In this section we present the comparison of the performances of the VOXES spectrometer with the ones obtained from the ray-tracing simulations. We used the XOP+SHADOW3 software \cite{Sanchez1,Sanchez2,Rebuffi1} implemented in the Oasys enviroment\cite{Oasys}. The XOP package includes all the codes that are necessary to evaluate the interaction of the X-rays with the optical components defined by the user.  In particular, we choose an extension, the \textit{SHADOWVUI} package that provides a Visual User Interface for the SHADOW ray-tracing program. 
The elements of the optical system are defined by objects that are called \textit{widgets} within a canvas that represents our workspace. Each widget contains the  user-defined parameters
by double-clicking on it and has to be connected with others by connectors. 
The \textit{Loop Point} widget allows to increase the statistics performing several runs of smaller size and can be used also to start the simulation from the source widget which we refer as \textit{target}.

\begin{figure}[h!]
    \centering
    \includegraphics[width=1.0\linewidth]
    {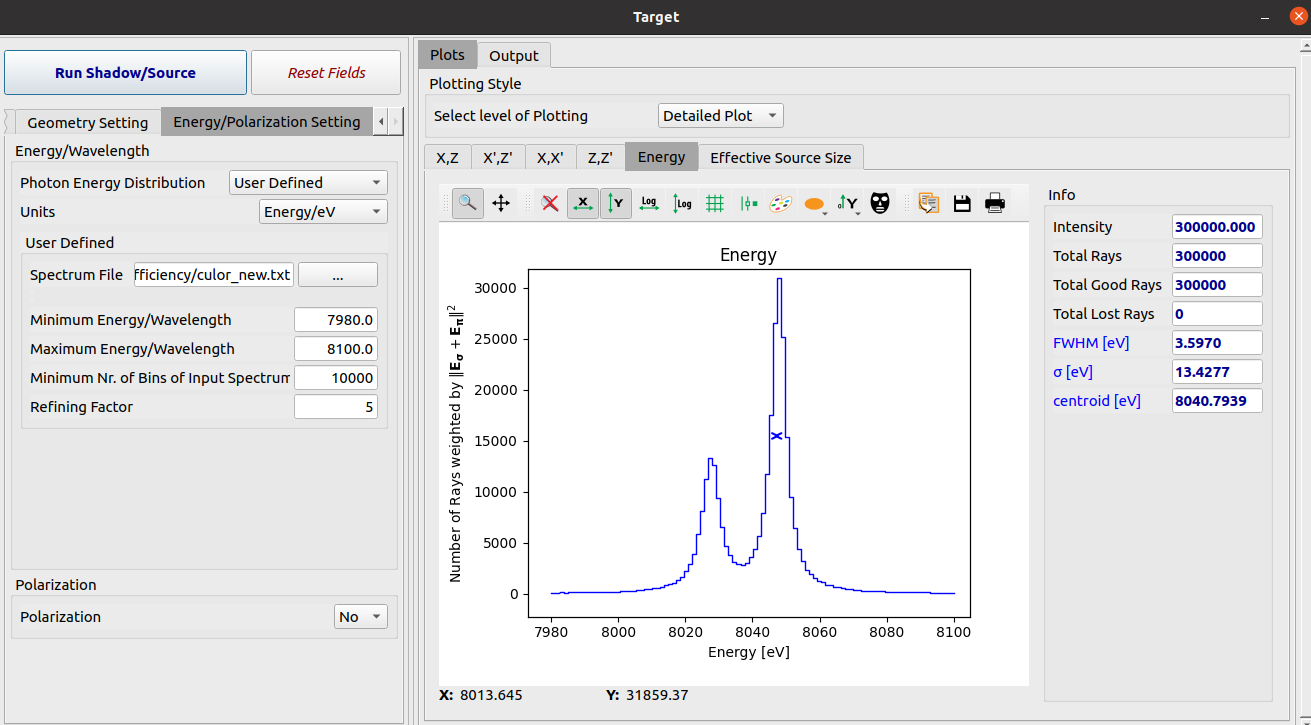}
    \caption{
 Snapshot of the \textit{target} widget where the energy spectrum given in input to the simulation is visible.
    }
    \label{snap_target}
\end{figure}

\begin{figure}[h!]
    \centering
    \includegraphics[width=1.0\linewidth]
    {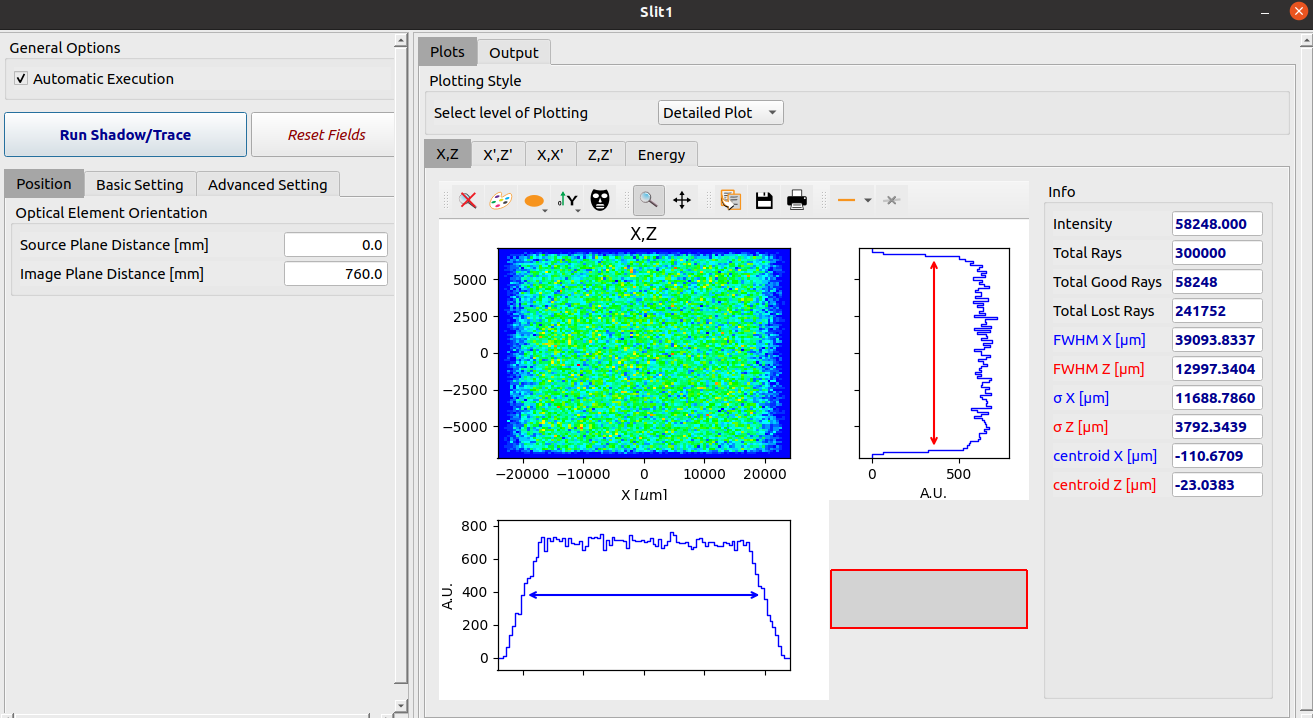}
    \caption{Snapshot of the slit widget used in the simulation of the copper Bragg spectra for  $\Delta\theta'$=0.7$^\circ$ and $S_0'$=1.1 mm; in the right part the setting of the position is shown while on the center, the two-dimensional XZ plot, together with the corresponding projections of the beam dimensions are visible.}
    \label{slit_snap}
\end{figure}

The \textit{target} is generated by means of Shadow Geometrical Source and we defined it as a $12 \times 1.3$ mm rectangular 0.1 mm thick to prevent any losses. 
In the same widget we provide as input the energy spectrum of the copper (or iron) X-rays. It is sampled with a double Lorentzian function, the widths of which are
obtained from the paper \cite{Holzer}. In the snapshot of the Fig.~\ref{snap_target} from the Geometrical Source widget, an example of the copper spectrum input is reported; on the left we inserted the energy range of the spectrum, the binning and the path of the input file while on the right the histogram statistics is listed.
After that, we defined other two widgets that include two python codes to take into account the target rotation of 45$^\circ$ and to consider the
X-ray divergence.

\begin{figure}[h!]
    \centering
     \includegraphics[width=1.0\linewidth] {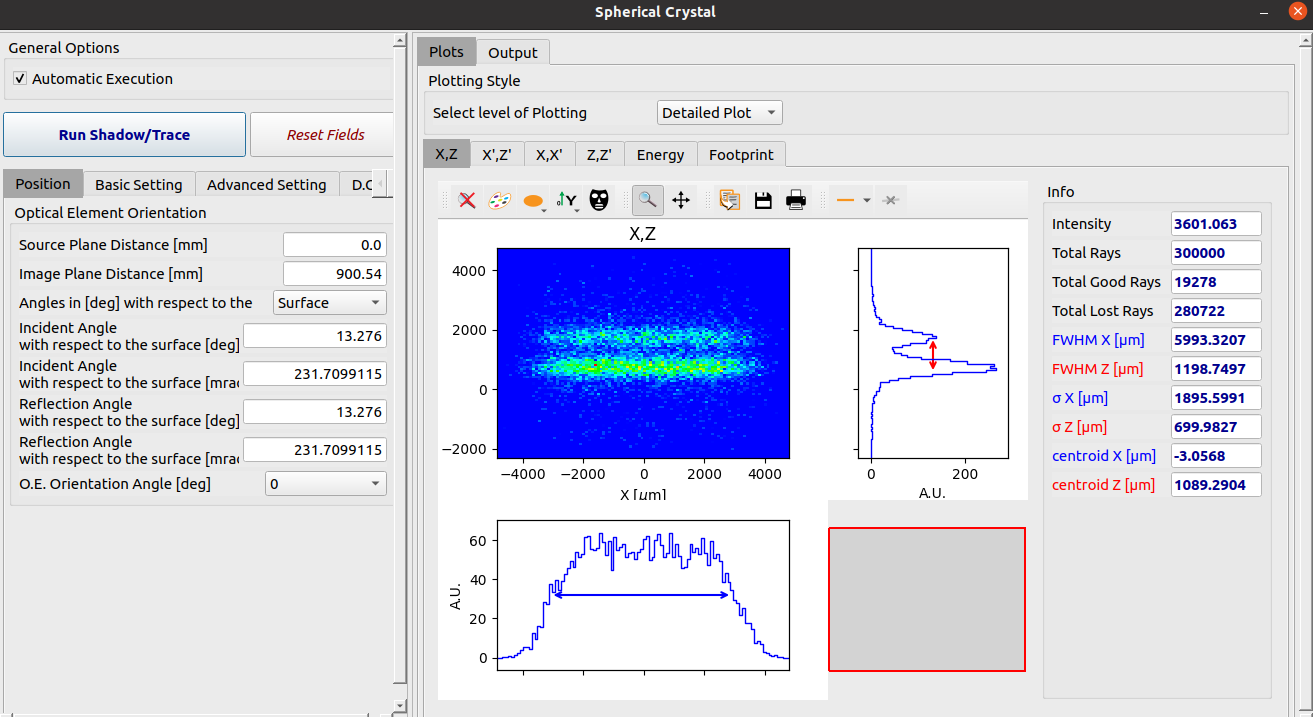}
    \caption{XZ plot of the XRF photons, emitted from the effective source and reflected by the  HAPG crystal, as seen from the MYTHEN2 detector point of view (top);
    intensity as a function of the Z-coordinate (bottom). 
    }
    \label{xz_sim}
\end{figure}

\begin{figure}[h!]
    \centering
    \includegraphics[width=0.7\linewidth]
     {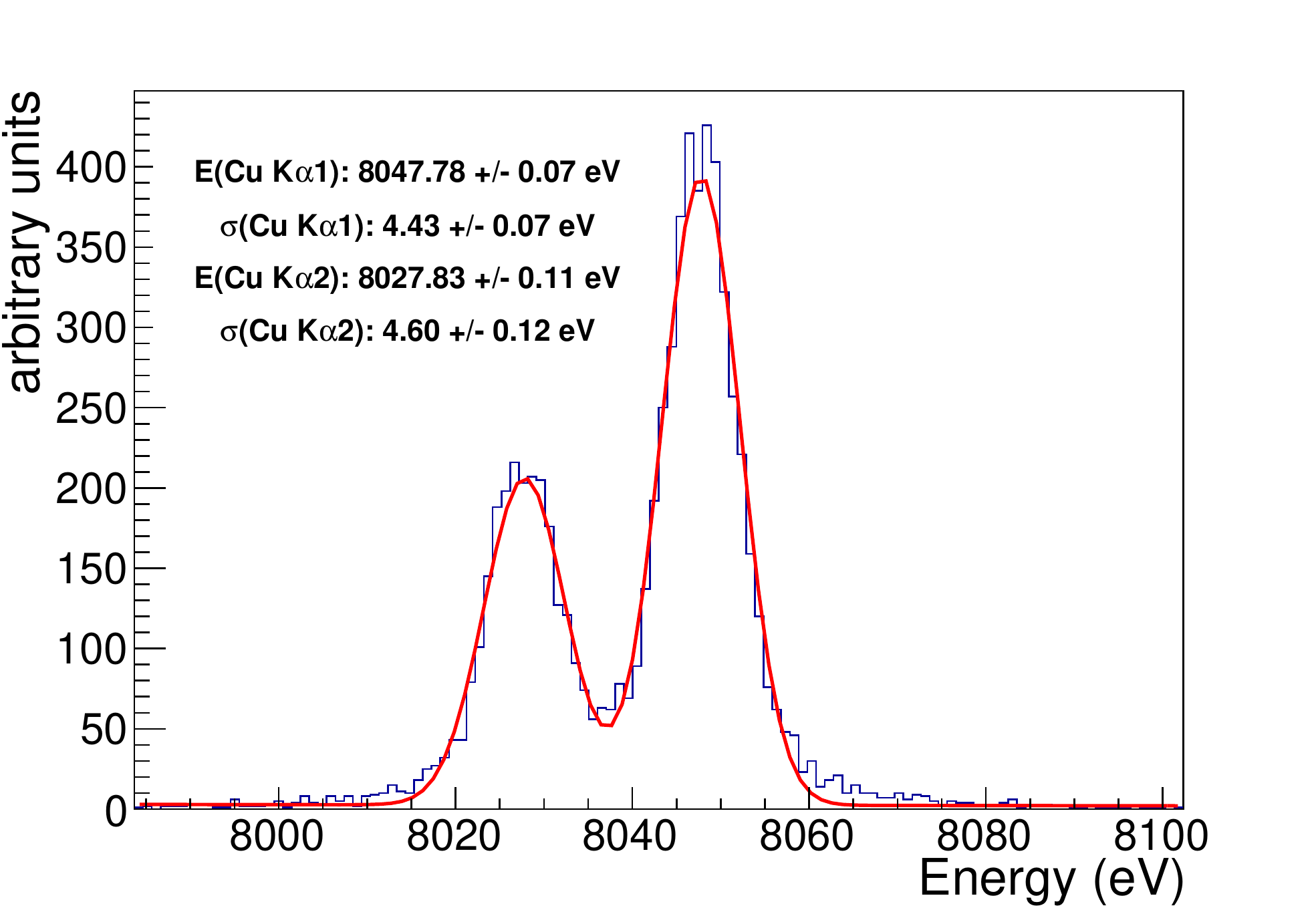}
     
    \caption{Simulated spectrum after energy conversion and fit with a gaussian function for copper,
    $\Delta\theta'$=0.7,$S_0'$=1100 mm
    }
    \label{calib_spectrum}
\end{figure}

\noindent Then, we implement the description of the box hole and the two slits system in order to be able to set their positions and apertures. These last two widgets have a very important role in the simulation
because they define the effective source size $S_0'$ and the angular divergence $\Delta\theta'$.
The snapshot of the Fig.~\ref{slit_snap} is referred to the widget of the first slit used in the simulation of the copper Bragg spectra for  $\Delta\theta'$=0.7$^\circ$ and $S_0'$=1.1 mm; in the right part the setting of the position are shown while on the center, the two-dimensional XZ plot, together with the corresponding projections of the beam dimensions, are reported. In these plots, it is possible to notice the effective source size of 1.1 mm. 
The following step consists in the simulation of the HAPG crystal by using the Shadow Spherical
Crystal widget. Here, we introduced the value of the curvature radius ($\rho_c$=206 mm), the mosaicity ($\omega$=0.1$^\circ$),
usually provided by the producer and also the position and the nominal Bragg angle at which the reflection occurs as shown in the snapshot reported in the Fig.~\ref{xz_sim} on the left column; on the center instead, the XZ plot of the XRF photons emitted from the effective source and reflected by the HAPG crystal is shown in case of a copper target $\Delta\theta'$=0.7$^\circ$ and $S_0'$=1.1 mm, together with the projections, in arbitrary units, of the Z and X coordinates on the right and on the bottom of the two-dimensional plot respectively. The Z-plot converted in energy and fitted with a gaussian function is shown in the Fig.~\ref{calib_spectrum} where it is possible to appreciate the good matching of the two XRF emission lines $K\alpha_{1,2}$ both in terms of energy and resolution with the experimental one of the Fig.~\ref{exp_spectrum}. The small difference between the resolution values is attributable to the MYTHEN2-1D cross talk between adjacent strips and to possible misalignment in the slits position.



\noindent To evaluate the reflection efficiencies from the simulations we used the following ratio \cite{nostro_JAAS}: 
\begin{equation}
\epsilon^{R, Sim}_{\Delta\theta',S_0'}=\frac{Intensity_{\Delta\theta',S_0'}}{Good\; rays_{\Delta\theta',S_0'}}  
\label{effsim_eq}
\end{equation}

\noindent where  $Good Rays_{\Delta\theta',S_0'}$ refers to the number of photons emitted by the source and reaching the MYTHEN2 detector, while $Intensity_{\Delta\theta',S_0'}$ accounts also for the HAPG crystal reflectivity. These quantities are displayed in the right part of the snapshot of the Fig.~\ref{xz_sim}.

\begin{figure}[h!]
\centering
  \includegraphics[width=0.6\linewidth]{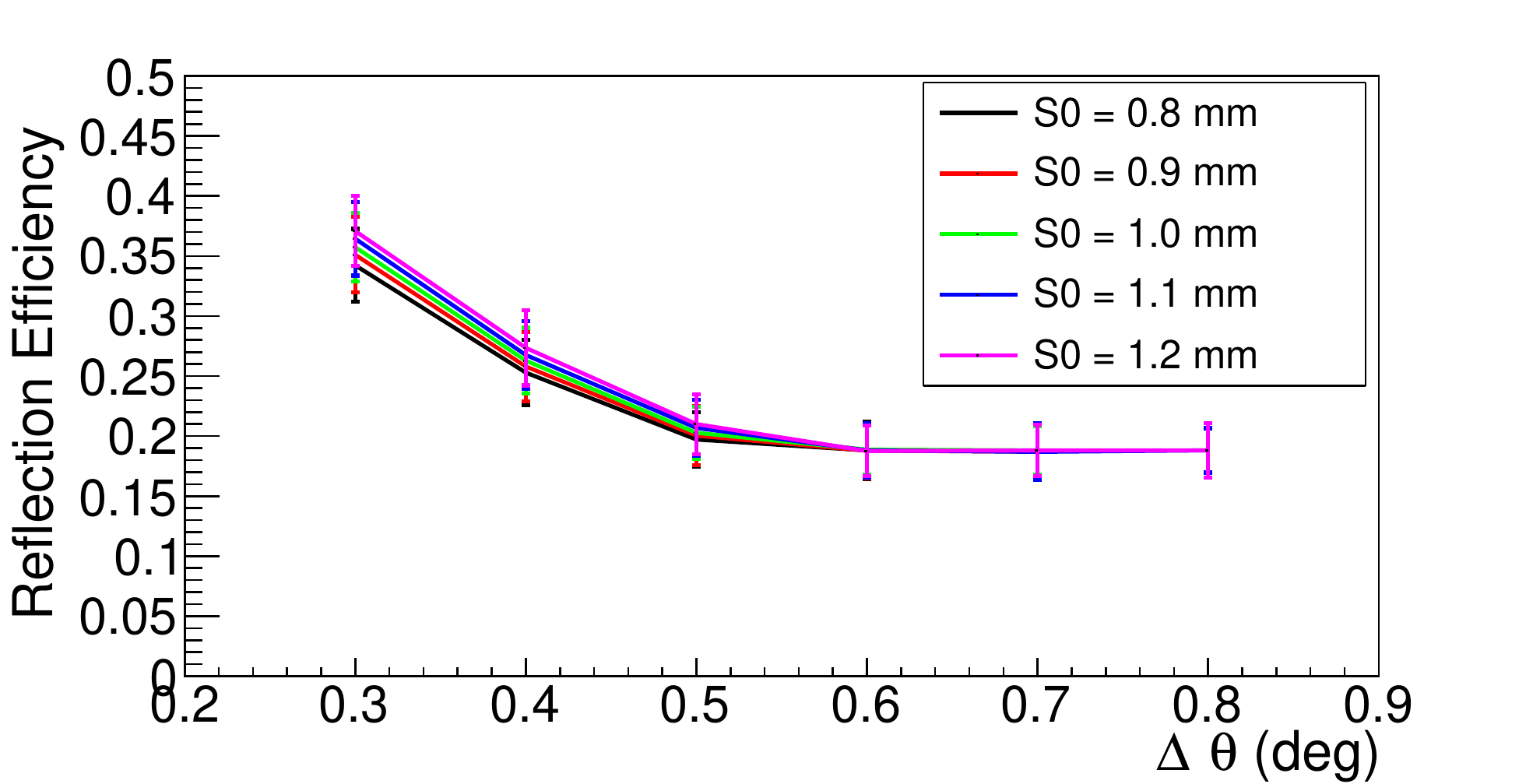}
   \includegraphics[width=0.6\linewidth]{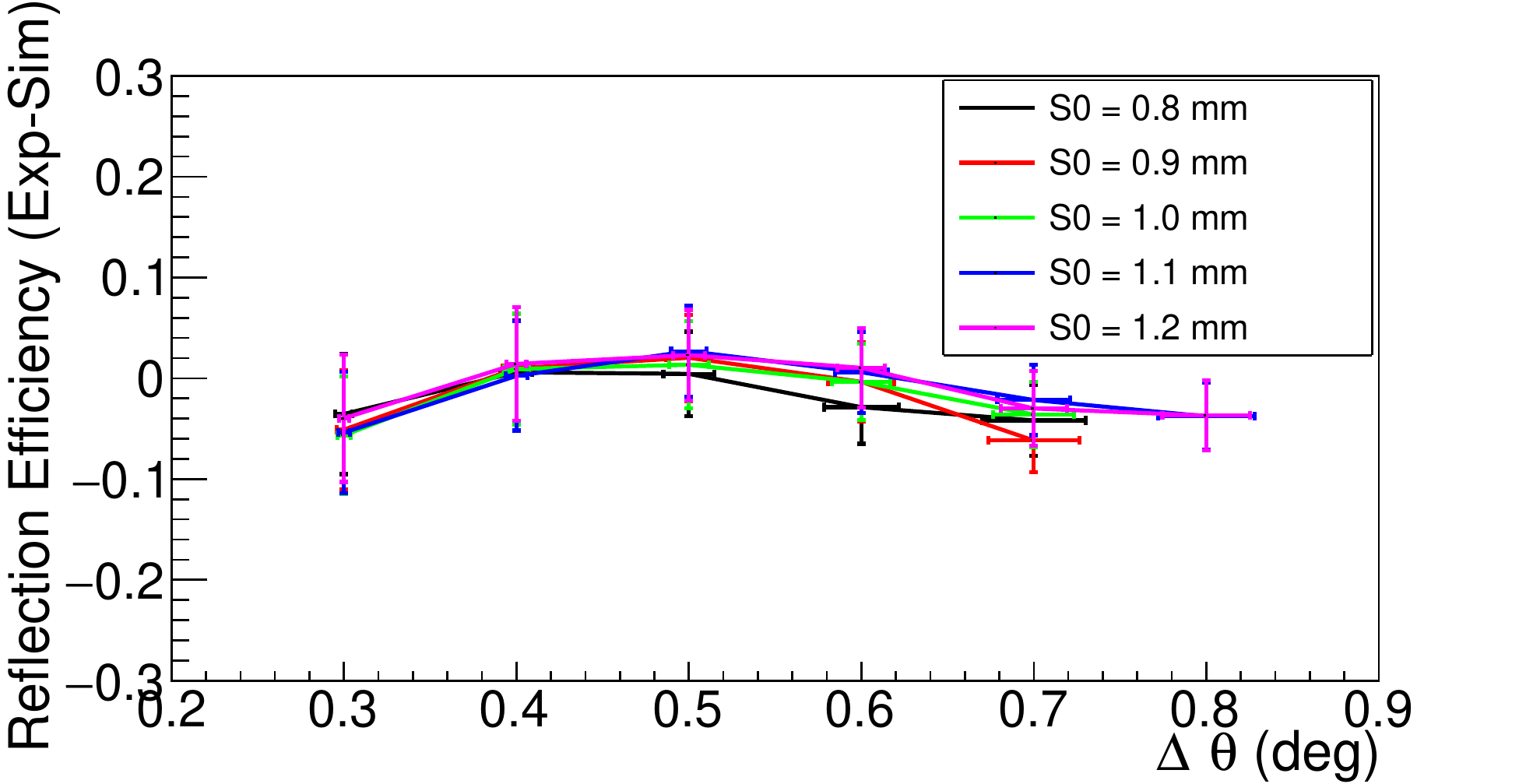}
  \caption{Reflection efficiency from simulations (top) and difference between the experimental and simulated reflection efficiency (bottom) for copper, ($\rho$=206mm) \cite{nostro_JAAS}.}
  \label{simeff_cu}
\end{figure}

\begin{figure}[h!]
\centering
 \includegraphics[width=0.6\linewidth]{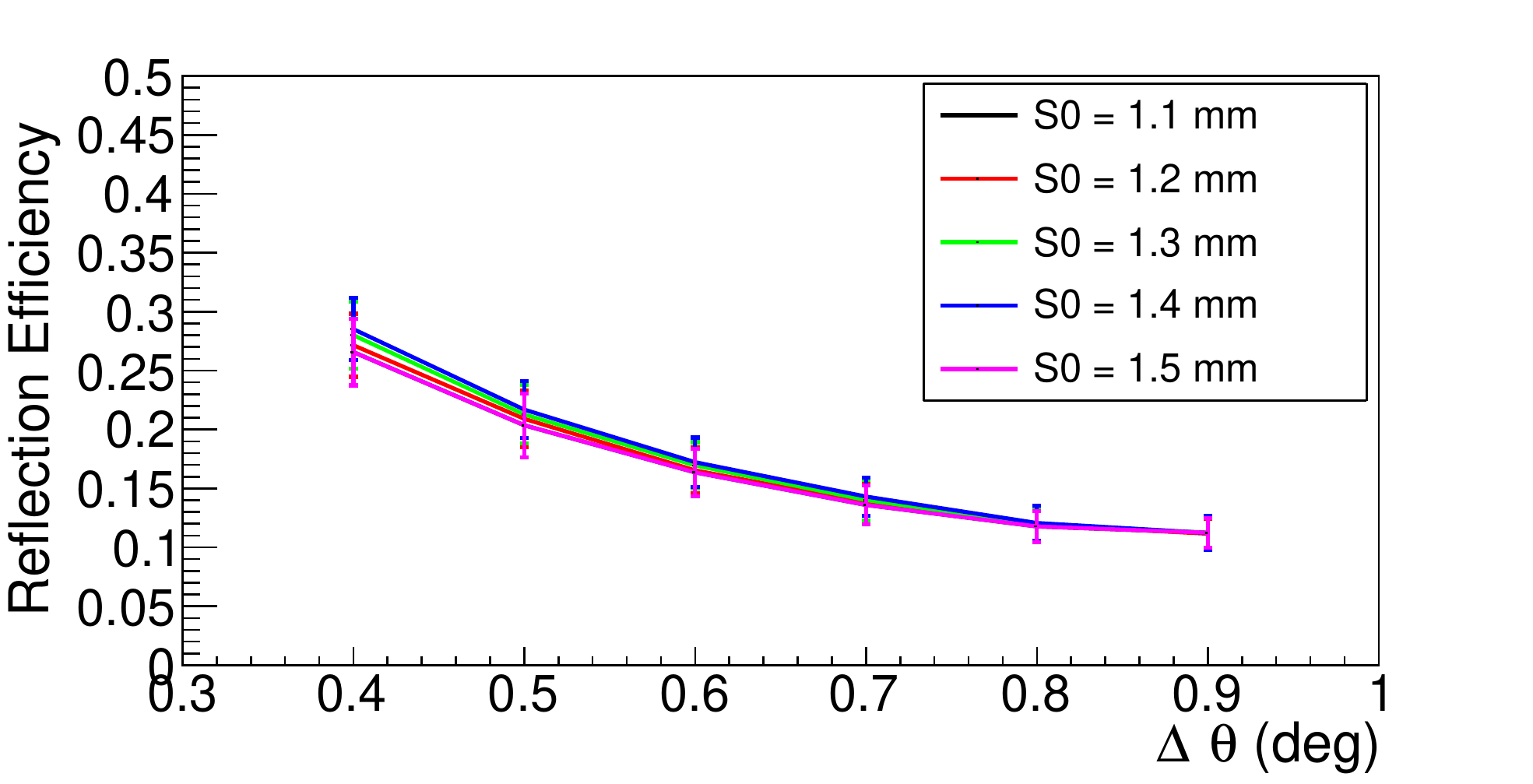}
   \includegraphics[width=0.6\linewidth]{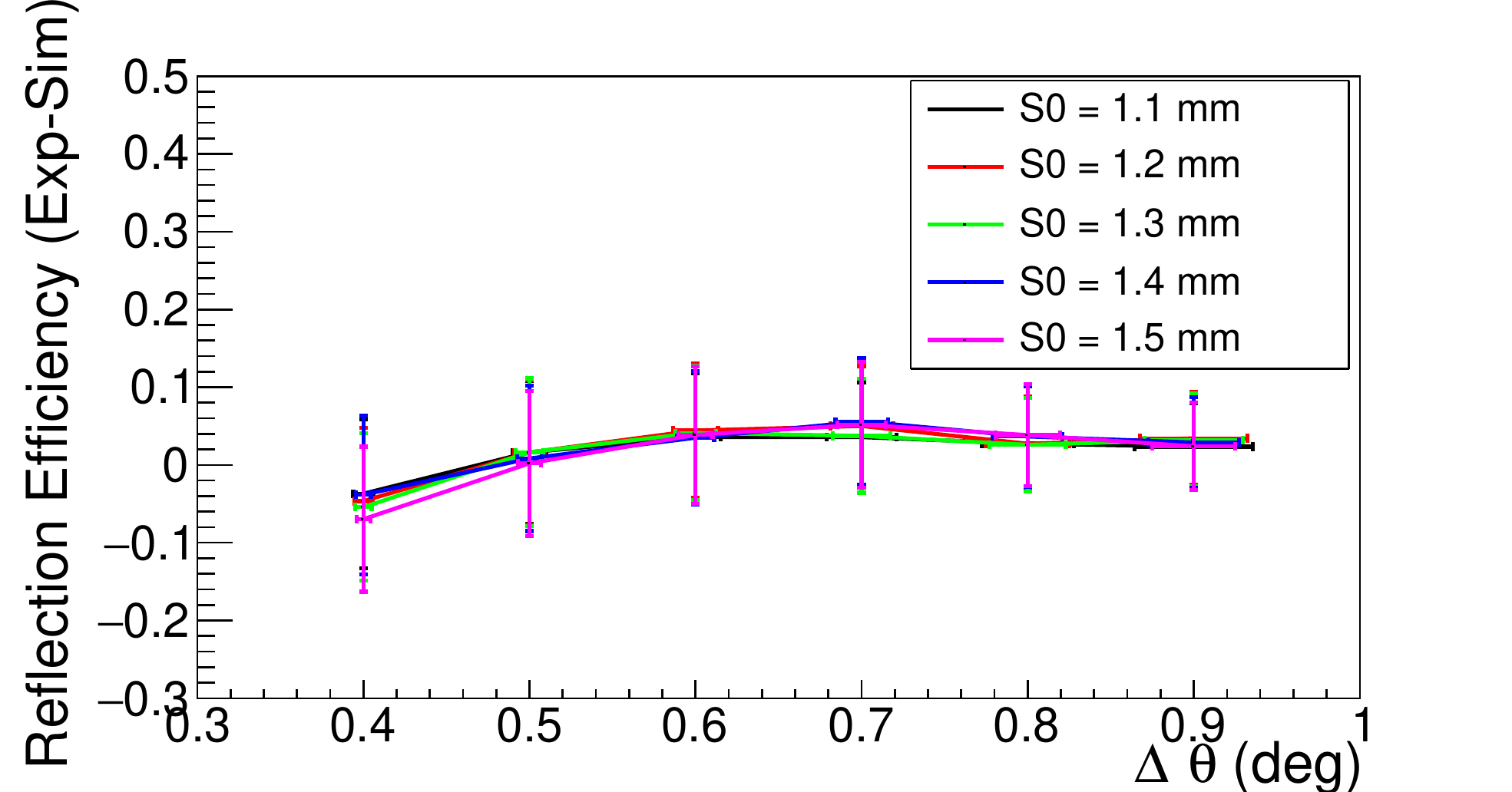}
  \caption{Simulated reflection efficiency (top) and difference between the experimental and simulated reflection efficiency (bottom) for iron, ($\rho$=206mm)  \cite{nostro_JAAS}.}
  \label{simeff_fe}
\end{figure}

The results obtained in terms of reflection efficiency for different effective source sizes, $S_0'$, as a function of the angular divergence $\Delta\theta'$ from simulated data are shown for copper and iron in Figs.~\ref{simeff_cu} and ~\ref{simeff_fe} respectively,  in the top pads, while the differences between the experimental and the simulated efficiency  are given in the bottom pads. 

\noindent The results highlight how the simulation well reproduces the experimental data within the associated errors.
For each point of the simulated reflection efficiencies plots, the error bars are the result of the sum in quadrature of the statistical (almost negligible) and systematic errors.
To evaluate the systematic uncertainty we performed simulations fixing the $S_0'$ and $\Delta\theta'$ values and varying different parameters as the position of the slits, crystal and MYTHEN detector ($\pm$5 mm), the opening of the slits ($\pm$0.05 mm), their possible misalignment in the x-coordinate ($\pm$5 mm) and in the z-coordinate ($\pm$0.1 mm) and finally the mosaicity ($\pm$0.01$^\circ$). 
\begin{figure}[h]
\centering
\includegraphics[width=0.6\linewidth]{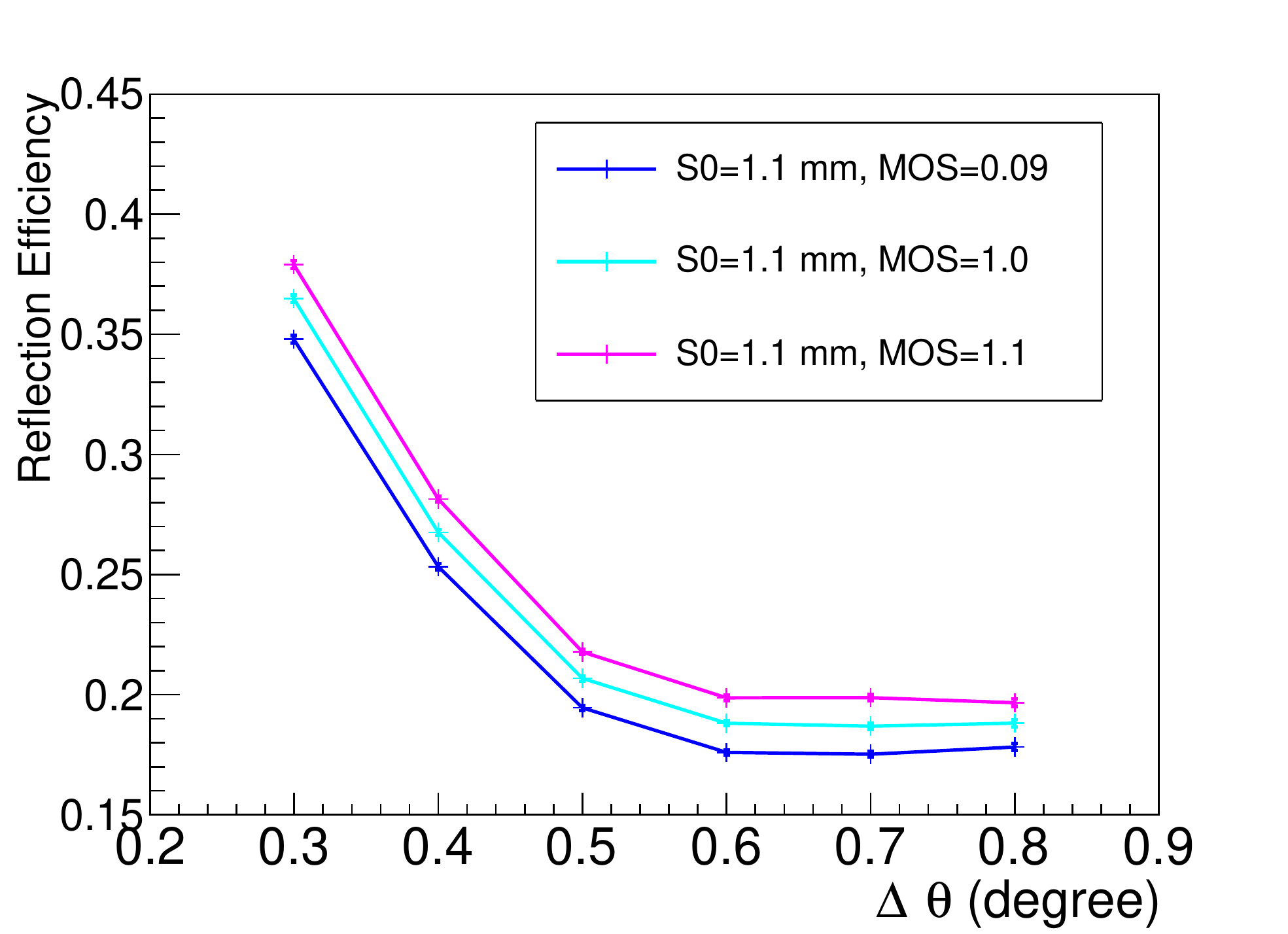}
  \includegraphics[width=0.6\linewidth]{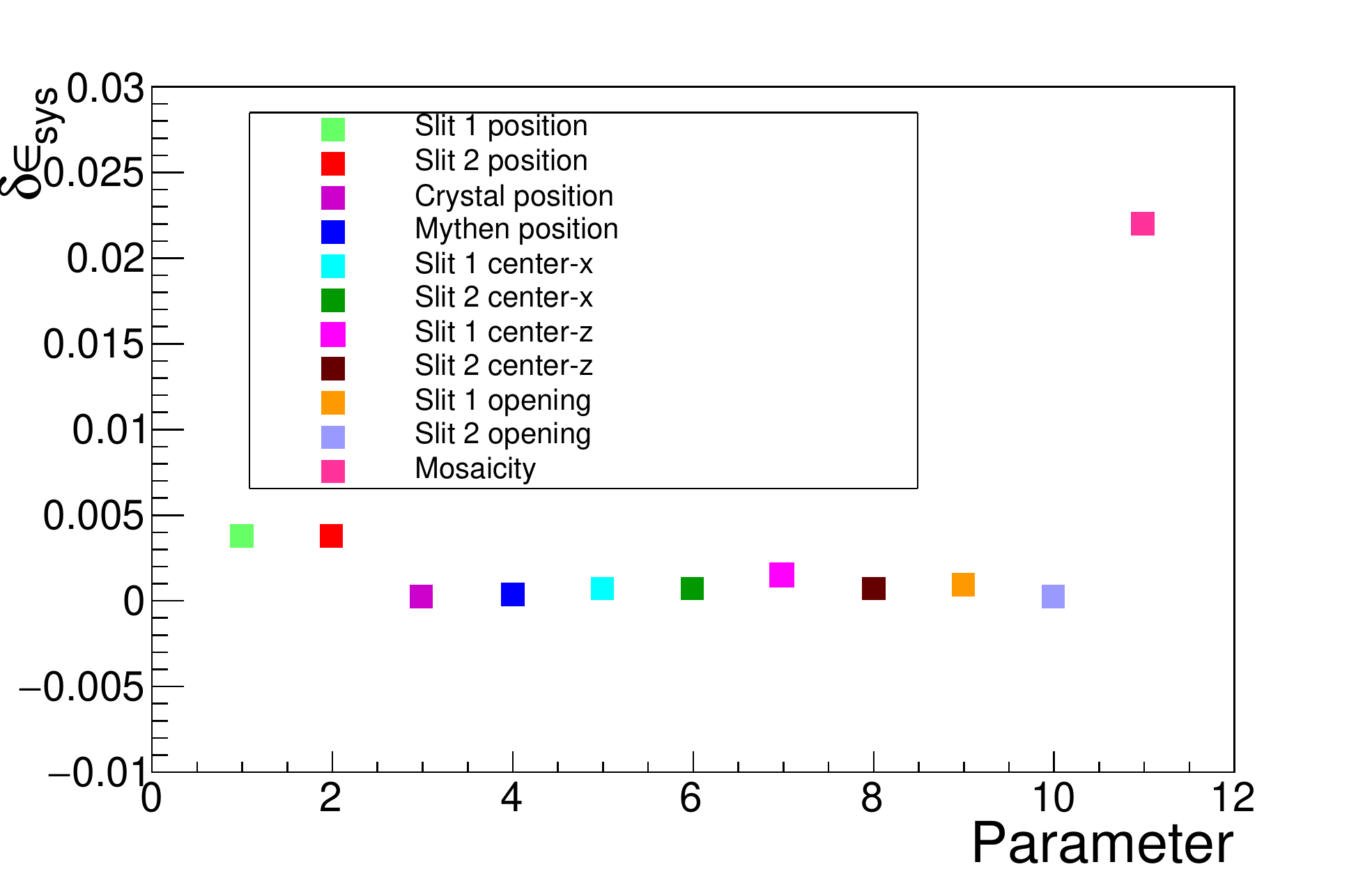}
  \caption{Reflection efficiencies obtained for mosaicity values 0.09, 0.1 and 0.11 are represented in blue, cyan and magenta respectively (top). Contributions to the systematic uncertainties for copper $S_0'=1100\mu$m and $\Delta\theta'=0.7^{\circ}$ (bottom).}
  \label{sys_err}
\end{figure}

The consequent variation of the reflection efficiency values has been accounted as systematic error.
In the Fig.~\ref{sys_err} we reported the study carried out in case of use of a copper target for $\Delta\theta'$=0.7$^\circ$, $S_0'$=1.1 mm. 
In particular, on top the reflection efficiencies obtained for mosaicity values 0.09, 0.1 and 0.11 are represented in blue, cyan and magenta, respectively, while on bottom the systematic uncertainties referred to all analyzed parameters  are summarized.  
 As one can see, most parameters gave a contribution 
 less than 5 per mill while the variation of the mosaicity resulted of the order of 2\%. This value has been taken as systematic uncertainty associated to the reflection efficiency.

\begin{figure}[h!]
\centering
   \includegraphics[width=0.6\linewidth]{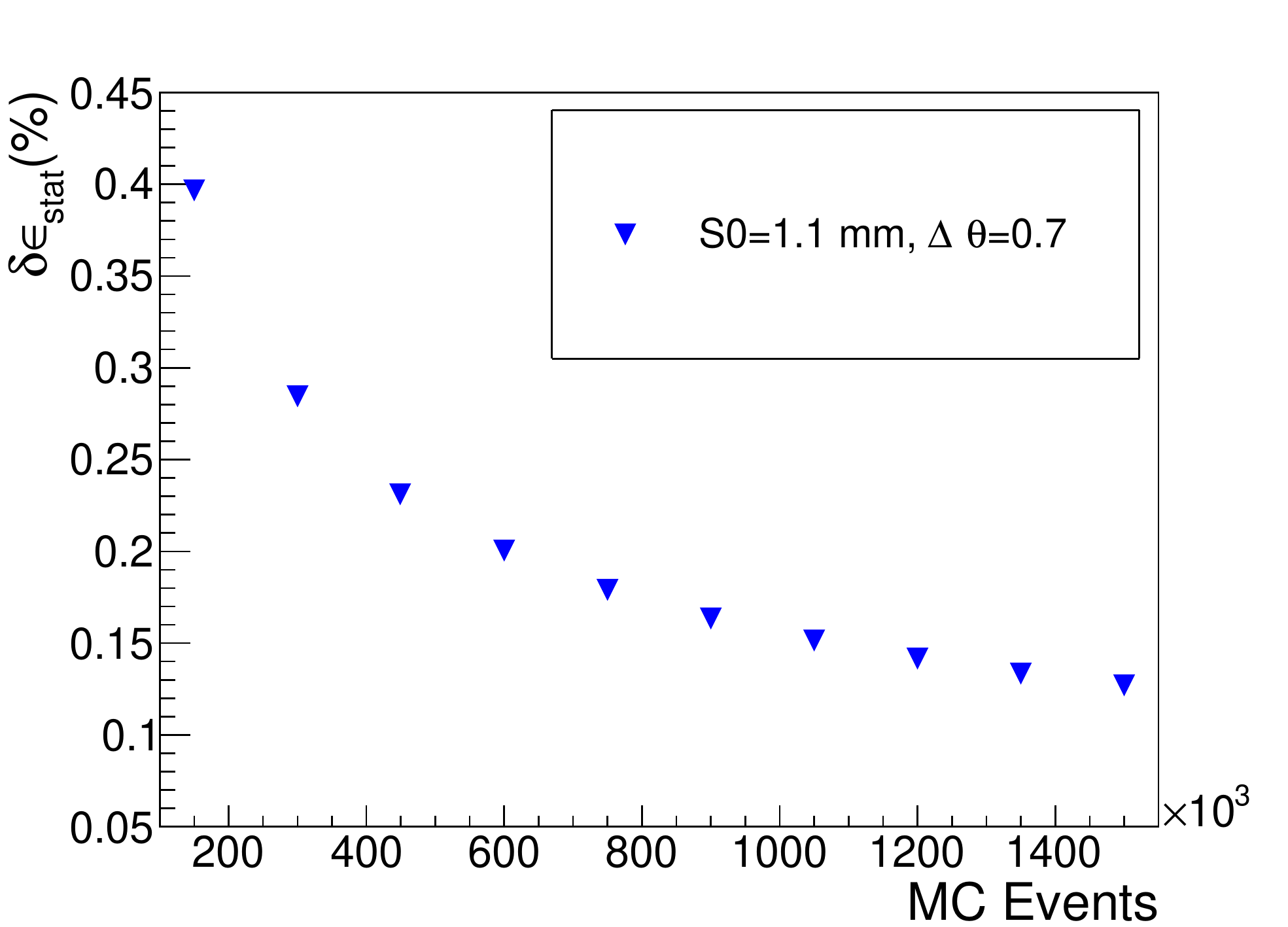}
  \caption{Variation of the statistical error on the reflection efficiency as a function of the number of generated events.}
  \label{stat_err}
\end{figure}

\noindent The number of events used as input in the simulations has been chosen according to the plot shown in Fig.~\ref{stat_err}, where the statistical error on the reflection efficiency ($\delta\epsilon_{stat}$) as a function of the MC simulated events, as example, for the $S_0'=1100\mu $m\,, and $\Delta\theta'=0.7^{\circ}$ copper case, is reported. In our final simulations, we used $1.5\times10^6$ events not to affect the precision of the efficiency simulations by low statistics. 

\section{Discussion}
 
In this work we presented the experimental procedure
followed to carry out the reflection efficiency measurements for $\rho$ = 206.7 mm cylindrically bent HAPG crystal using $CuK\alpha_{1,2}$ and $FeK\alpha{1,2}$ XRF lines. 
The results highlight that the measured reflection efficiencies appear almost unaffected by the source size while show a slight dependence on the beam divergence.

\noindent The simulation software and each step of the simulations performed in order to check the consistency with the experimental data were also described.
The response of the spectrometer has been validated by the ray tracing simulations, both in terms of the spectra reproduction and of the reflection efficiency measurement and a good agreement within the errors was found. Consistent and reliable simulated data are crucial to get information on the efficiency of the spectrometer which is fundamental for the experimental apparatus for the proposal of future experiments.

\begin{center}
\small
\section*{Acknowledgments}
\end{center}
This work was supported by the 5th National Scientific Committee of INFN in the framework of the Young Researcher Grant 2015, n. 17367/2015. This project has received funding from the European Union’s Horizon 2020 research and innovation program EU STRONG-2020, under grant agreement No 824093.
We thank the LNF and SMI staff, in particular the LNF SPCM service
and Doris Pristauz-Telsnigg, for the support in the preparation
of the setup.


\begin{thebibliography}{30}



\bibitem{Barn}
R. Barnsley, et al., Versatile high resolution crystal spectrometer with x-ray charge coupled device detector.
Rev. Sci. Instrum., 2003, \textbf{74}, 2388.

\bibitem{Legall}
H. Legall et al., A new generation of X-ray optics based on pyrolitic graphite, Proceedings of FEL 2006, 2006, 798.

\bibitem{Grigorieva}
I. G. Grigorieva and A. A. Antonov, 
HOPG as powerful x-ray optics, 
X-Ray Spectrometry, 2003, 32. 






\bibitem{Scordo_Cond}
A. Scordo et al., Pyrolitic Graphite Mosaic Crystal Thickness and Mosaicity Optimization for an Extended Source Von
Hamos X-ray Spectrometer, Condens. Matter, 2019,\textbf{4}, 38.



\bibitem{Scordo_JAAS}
A. Scordo et al., High resolution multielement XRF spectroscopy of
extended and diffused sources with a graphite mosaic crystal based Von Hamos spectrometer., J. Anal. At. Spectrom., 2020, \textbf{35},
155-168.

\bibitem{nostro_JAAS}
 A. Scordo, V. De Leo, C. Curceanu, M. Miliucci and F. Sirghi, Efficiency measurements and simulations of a HAPG based Von Hamos spectrometer for large sources, J. Anal. At. Spectrom.,2021.





\bibitem{CXRO}
B. Henke, E. Gullikson and J. Davis,
Atomic Data and Nuclear
Data Tables, 1993, \textbf{54}, 181-342.


\bibitem{Sanchez1}
M. Sanchez del Rio et al., A conceptual model for ray tracing calculations with mosaic crystals, Review of Scientific Instruments, 1992, \textbf{63}, 932-935.

\bibitem{Sanchez2}
M. Sanchez del Rio et al., SHADOW3: a new version of the synchrotron X-ray optics modelling package, Journal of Synchrotron Radiation, 2011, \textbf{18}, 708-716.

\bibitem{Rebuffi1}
L. Rebuffi and M.  Sanchez del Rio, ShadowOui: a new visual environment for X-rayoptics and synchrotron beamline simulations, Journal of Synchrotron Radiation, 2016, \textbf{23}, 1357-1367.


\bibitem{Oasys}
L. Rebuffi and M.  Sanchez del Rio, OASYS (OrAnge SYnchrotron
Suite): an open-source graphical environment for
x-ray virtual experiments, 2017, 28.

\bibitem{Holzer}
G. Hölzer et al., $K\alpha_{1,2}$ and $K\beta_{1,3}$ x-ray emission lines of the 3d transition metals, Phys. Rev. A, 1997, \textbf{56}, 4554-4568.

\end{thebibliography}
\end{document}